# Матрица проводимости мультиконтактных полупроводниковых структур с краевыми каналами

Э.Ю. Даниловский[1,*], Н.Т. Баграев[1,2]

[1] Физико-технический институт им. А.Ф. Иоффе Российской академии наук, 194021 Санкт-Петербург, Россия.

[2] Санкт-Петербургский государственный политехнический университет, 195251 Санкт-Петербург, Россия

* e-mail: danilovskii@mail.ioffe.ru

**Conductance matrix of multi terminal semiconductor devices with edge channels**

E.Yu. Danilovskii[1] and N.T. Bagraev[1,2]

[1]Ioffe Physicotechnical Institute, Russian Academy of Sciences,

194021 St. Petersburg, Russia

[2]St.Petersburg State Polytechnical University, 195251 St.Petersburg, Russia

**Abstract** This paper presents a method for determining the conductance matrix of multi terminal semiconductor structures with edge channels.In order to identify the conductance matrix, we suggest to measure the values of the voltages $U_{i,j}$ between all probes at different directions of the highly-stabilized current $I_{i,j}$, where $i,j$ – indexes of probes. Then, the data obtained have to be used for the solution of the system of the linear algebraic equations based on the Kirchhoff's circuit laws for each from the eight terminals. Thus, the conductance matrix, **G**, is followed to be created taking account of the instrument and statistical accuracy of every element $G_{i,j}$. This method appears to be applied within frameworks of the ordinary Landauer-Buttiker formalism for the carrier transport analysis in the regime of both the quantum Hall Effect and the quantum spin Hall Effect. The proposed method proves to take into account principally the contribution of the probes resistance in the formation of the matrix conductance elements. Finally, the possibilities of the practical application of this method to develop new versions of analog cryptographic devices are discussed.

# Аннотация

В работе предложен метод определения матрицы проводимости мультиконтактных полупроводниковых структур с краевыми каналами. В основе метода лежит решение системы линейных алгебраических уравнений (СЛАУ) на базе уравнений Кирхгофа, составленных из разностей потенциалов $U_{ij}$, измеренных при стабилизации токов $I_{kl}$, где $i, j, k, l$ - номера контактов. Матрица, полученная в результате решения СЛАУ, полностью описывает исследуемую структуру, отражая её геометрию и однородность. Данный метод может найти широкое применение при использовании известного формализма Ландауэра-Буттикера для анализа транспорта носителей в режимах квантового эффекта Холла и квантового спинового эффекта Холла. В рамках предлагаемого метода учитывается вклад сопротивления контактных площадок, $Rc$, в формирование элементов матрицы проводимости. Рассматриваются возможности практического применения полученных результатов при разработке аналоговых криптографических устройств.

# 1. Введение

Одним из актуальных направлений в современной физике конденсированного состояния является изучение квантованной проводимости полупроводниковых наноструктур, демонстрирующих в сильном магнитом моле наличие краевых каналов, возникновение которых лежит в основе квантового эффекта Холла (КЭХ) [1, 2]. Отдельный интерес представляют топологические парные краевые каналы, в которых носители имеют противоположную ориентацию спинов в отсутствие внешнего магнитного поля, что открывает возможности для регистрации квантового спинового эффекта Холла (КСЭХ) [3-7]. Исследования в этом направлении могут найти широкое практическое применение в спинтронике и разработке квантовых компьютеров на базе новых материалов - топологических изоляторов и сверхпроводников [8, 9].

При описании явления краевой проводимости в режиме КЭХ и КСЭХ для расчета сопротивления образцов, имеющих несколько контактов, хорошо зарекомендовал себя формализм Ландауэра-Буттикера [10, 11]. В рамках этого подхода полный ток, протекающий через образец, в матричной форме записывается как $I = G \cdot V$, где $I$ и $V$ – столбцы токов и напряжений для каждого из $N$ контактов, $G$ - матрица проводимости размерности $N \times N$.

Цель настоящей работы заключалась в разработке метода регистрации матрицы проводимости мультиконтактных полупроводниковых структур и демонстрации возможности его применения на примере восьми контактного холловского мостика в режимах КЭХ и КСЭХ. Для того чтобы определить матрицу проводимости в рамках предложенного метода, необходимо измерить падения напряжений $U_{i,j}$ между каждой парой контактов при условии стабилизации тока $I_{k,l}$, где $i,j,k,l$ – номера контактов. Индексы ($k$, $l$) подбираются так, чтобы из полученных данных становилось возможным для *каждого* из восьми контактов составить систему линейных алгебраических уравнений, базирующихся на правилах Кирхгофа. Решение полученных уравнений позволяет получить матрицу проводимости $G$, а также рассчитать с учетом статистической и приборной погрешности измерений $U_{i,j}$ точность определения матричных элементов $G_{i,j}$. Однако экспериментальная регистрация матрицы проводимости полупроводниковой мультиконтактной структуры с краевыми каналами представляет известную сложность, поскольку необходимо принимать во внимание ряд побочных эффектов. В частности, возникают дополнительные эдс, вследствие наличия конечного сопротивления контактных площадок и неполной диссипации энергии баллистических носителей на контактах структуры. В рамках предложенного метода детально исследован вклад сопротивления контактных площадок, $Rc$, в формирование элементов матрицы проводимости в режимах КЭХ и КСЭХ. Рассмотрены возможности практического применения полученных результатов при разработке аналоговых криптографических устройств.

**2. Результаты и обсуждение**

**2.1. Измерение матрицы проводимости мультиконтактной полупроводниковой структуры**

Рассмотрим алгоритм построения матрицы проводимости на примере восьми контактного устройства (см. рис. 1). В этом случае матрица содержит 8 x 8 = 64 элемента, из них 8 – диагональных, каждый из которых равен сумме элементов в соответствующей строке, взятой со знаком «-», и 64 – 8 = 56 недиагональных элементов, для нахождения которых необходимо составить систему из 56 линейных алгебраических уравнений (СЛАУ). Для этого требуется записать уравнения Кирхгофа для каждого из восьми контактов, посредством измерения всех возможных разностей потенциалов $U_{ij}$, $i,j = 1 – 8$, $i \neq j$, при условии стабилизации тока для семи различных пар токовых контактов. Величину стабилизированного тока рекомендуется подбирать так, чтобы выделяемая мощность джоулевого тепла была много меньше 1мВт [12]. При этом основным критерием подбора служит баланс между выделением образцом тепла и его поглощением теплоносителем, используемым для поддержания постоянной температуры. Таким образом, в результате измерений получаются 8 СЛАУ из семи уравнений вида:

$$I_j = \sum_{i=1, i \neq j}^{8} G_{ij} \cdot U_{ij}, \qquad (1)$$

где $I_j$ — ток, протекающий через контакт с номером $j$, $G_{ij}$ — искомые матричные элементы матрицы проводимости. При этом выбор пар токовых контактов $(k,l)$ должен быть обусловлен симметрией изучаемого мультиконтактного устройства. Так, в случае холловской геометрии контактов (см. рис. 1), такими парами будут: $(k,l)$ = (1,5); (1,4); (1,3); (1,2); (2,6); (2,7); (2,8), т.к. среди такого набора нет двух пар, эквивалентных с точки зрения симметрии. В противном случае, (например, если в наборе пар токовых контактов есть (1,4) и (1,6)) матрица каждой СЛАУ будет содержать две строки с элементами, отличающимися попарно друг от друга лишь в пределах погрешности измерения. В результате чего полученная матрица проводимости либо будет содержать заведомо ложные решения, либо в принципе не сможет быть построена ввиду равенства нулю определителя СЛАУ (1).

Для сокращения времени измерений целесообразно при стабилизации тока измерять разности потенциалов только между 8 различными парами контактов из 28 возможных. Оставшиеся 20 значений разностей потенциалов, необходимых для составления СЛАУ, можно вычислить, воспользовавшись свойством аддитивности потенциала электрического поля.

Для нахождения погрешности полученных матричных элементов $G_{ij}$, необходимо рассчитать статистическую и приборную ошибки для разностей потенциалов $U_{ij}$, а затем, с помощью известных формул для погрешностей суммы и произведения нескольких величин [13], вычислить погрешности для определителей матрицы системы уравнений (1) и всех её миноров. Далее, воспользовавшись методом Крамера, становится возможным получить погрешности для каждого из элементов обратной матрицы СЛАУ (1), которые после умножения на соответствующий столбец токов, дают искомые величины $\Delta G_{ij}$.

Важной особенностью описанного выше метода получения матрицы проводимости является независимость определения матричных элементов, симметричных относительно диагонали, $G_{ij}$ и $G_{ji}$, и отвечающих противоположным направлениям протекания тока. В случае, когда матричные элементы обладают свойством взаимности, т.е. $G_{ij} = G_{ji}$, матрица проводимости является, по сути, эквивалентной схемой исследуемого мультиконтактного устройства и полностью его описывает при заданных условиях эксперимента (величине стабилизированного тока, температуры, давления), подобно «белому ящику», описанному Норбертом Винером [14]. Более того, из анализа величин матричных элементов, соответствующих симметрии геометрии исследуемого образца, можно сделать вывод о степени его однородности. Так, например, для планарной структуры, выполненной в геометрии восьми контактного холловского мостика, в случае равномерного легирования и равенства расстояний между продольными контактами, будут выполняться равенства $G_{23} = G_{34} = G_{67} = G_{78}$, $G_{24} = G_{68}$, $G_{12} = G_{81} = G_{45} = G_{56}$ и т.д.

Такое описание, однако, становится невозможным, если матричные элементы $G_{ij}$ не обладают свойством взаимности, $G_{ij} \neq G_{ji}$. Эти условия нарушаются, в частности, при наличии внешнего магнитного поля [15]. При этом симметрия по отношению к обращению времени требует выполнения соотношения $[G_{ij}]_{+B} = [G_{ji}]_{-B}$ [16]. Отдельный интерес представляет случай квантующих полей. Тогда матрица проводимости становится эквивалентна по своему смыслу матрице функции пропускания, полученной в рамках формализма Ландауэра-Буттикера, и по виду матрицы становится возможным сделать вывод о реализации краевого транспорта в режимах целочисленного квантового эффекта Холла и квантового спинового эффекта Холла [1, 16]. На данный момент существует целый ряд оригинальных работ, посвященных проблеме экспериментального доказательства наличия узкого проводящего слоя в 3D и 2D топологических изоляторах, а также попыткам визуализации этого слоя [8, 17]. Предложенный метод измерения матрицы проводимости позволяет не только решить эту проблему, но и связать экспериментальные данные с теоретическими результатами, полученными на базе формализма Ландауэра-Буттикера и матриц рассеяния при описании интерференционных квантовых поправок к проводимости [18-21]. В частности, матричные элементы могут отражать осцилляции Ааронова-Кашера продольной проводимости, которые возникают вследствие изменений величины спин-орбитального взаимодействия Рашбы в валентной зоне квантовой ямы в условиях приложения напряжения вертикального затвора [7, 20, 22-26]. Метод измерения матрицы проводимости также может быть распространен на случай кольцевых мультиконтактных наноструктур [20, 26]. При этом отдельный интерес представляет исследование зависимостей матричных элементов от величины индукции внешнего магнитного поля в режиме регистрации эффекта Ааронова-Бома [27-30].

При построении матрицы проводимости необходимо учитывать ряд побочных эффектов, приводящих к появлению дополнительных эдс на контактах, в результате чего могут возникать значительные ошибки при определении величин матричных элементов $G_{ij}$. К таким побочным эффектам относятся:

1) эдс асимметрии поперечных зондов [31]. Рассогласование в расстояниях между контактами может привести к весьма существенным искажениям величин матричных элементов, особенно в устройствах наноэлектроники. В зависимости от технологии изготовления структуры, точность $b$, с которой выставляется расстояние между контактами, $L$, может варьироваться от 10 нм до 1 мкм. Так для современной фотолитографии достигнута точность в 100 нм. Таким образом, относительная ошибка матричных элементов может достигать $a=b/L$.

2) Эффекты термоэдс, возникающие на контактах при наличии градиента температуры в исследуемой структуре, а также на узловых точках измерительной цепи в местах соединения двух различных металлов за счет эффекта Зеебека. Для минимизации вклада термоэдс рекомендуется использовать при монтаже измерительной цепи только соединения типа медь-медь, избегая при этом образования окислов, и строго выдерживать условие термостабилизации в процессе измерений. При низкотемпературных измерениях также можно проводить усреднение измеренных разностей потенциалов по двум направлениям стабилизированного тока.

3) эдс, возникающие вследствие наличия конечного сопротивления у контактных площадок, $Rc$. Так как для определения матрицы проводимости невозможно использовать только 4-х контактные схемы измерении, то для каждой пары токовых контактов ($k,l$) разности потенциалов $U_{kj}$ и $U_{ik}$ будут включать в себя паразитные падения напряжений $Rc \cdot I_{kl}$, что при расчете отразится на величинах всех матричных элементов $G_{ij}$. Чтобы свести эти ошибки к минимуму, необходимо подбирать геометрические размеры контактных площадок так, чтобы их сопротивления были минимальными, а также использовать только качественные омические контакты. Еще одним следствием наличия конечного сопротивления $Rc$, является усиление амплитуды шумов, сопутствующих измерениям [31]. В предложенной схеме выбора токовых контактов ($k,l$) = (1,5); (1,4); (1,3); (1,2); (2,6); (2,7); (2,8), контакты № 1 и №2 задействованы наиболее часто, вследствие чего погрешность определения матричных элементов, $\Delta G_{ij}$, для первой и второй строк матрицы проводимости будут заметно превышать погрешности всех остальных элементов.

4) Чистота и качество поверхности исследуемой структуры также играет существенную роль в формировании элементов матрицы проводимости,

в частности, вследствие эффектов шунтирования контактов поверхностными состояниями и оборванными связями.

## 2.2. Матрица проводимости в режиме квантового эффекта Холла

Все рассмотренные выше рассуждения могут быть применены при анализе поведения сопротивления мультиконтактных устройств в режиме целочисленного квантового эффекта Холла (ЦКЭХ) в рамках формализма Ландауэра – Буттикера (ФЛБ) [1, 16].

В этом случае ток, протекающий через контакт $j$, может быть выражен через напряжения на всех остальных контактах:

$$I_j = \sum_i G_{ij}[V_j - V_i], \tag{2}$$

$$G_{ji} \equiv \frac{2e^2}{h}\bar{T}_{j\leftarrow i} \tag{2'}$$

где $G_{ji}$ - проводимость канала из контакта $i$ в контакт $j$, которая удовлетворяет следующим условиям $\sum_i G_{ij} = \sum_i G_{ji}$ и $[G_{ij}]_{+B} = [G_{ji}]_{-B}$, $T_{j\leftarrow i}$ – обобщенная функция пропускания, которая учитывает число краевых каналов между контактами $j$ и $i$, а также вероятность рассеяния носителей в процессе движения. Первое условие отражает тот факт, что если напряжение на всех контактах одинаково, то протекающий ток равен нулю, а второе является следствием симметрии по отношению к обращению времени [16].

В режиме ЦКЭХ матричные элементы $G_{ij}$ не равны нулю только для каналов, соединяющих соседние контакты (1←8), (2←1), (3←2), (4←3), (5←4), (6←5), (7←6) и (8←7) восьми контактного холловского мостика (см. рис 2$a$). В простейшем случае, когда все ненулевые матричные элементы равны между собой и отвечают пропусканию системы из $M$ краевых каналов, $G_{ij} = M \cdot e^2/h = M \cdot G_c \equiv G_0$, выражение (2) может быть записано в матричной форме следующим образом (напряжение на контакте № 1 выбрано в качестве точки отсчета, $V_1 = 0$):

$$\begin{Bmatrix} I_2 \\ I_3 \\ I_4 \\ I_5 \\ I_6 \\ I_7 \\ I_8 \end{Bmatrix} = \begin{bmatrix} G_0 & 0 & 0 & 0 & 0 & 0 & 0 \\ -G_0 & G_0 & 0 & 0 & 0 & 0 & 0 \\ 0 & -G_0 & G_0 & 0 & 0 & 0 & 0 \\ 0 & 0 & -G_0 & G_0 & 0 & 0 & 0 \\ 0 & 0 & 0 & -G_0 & G_0 & 0 & 0 \\ 0 & 0 & 0 & 0 & -G_0 & G_0 & 0 \\ 0 & 0 & 0 & 0 & 0 & -G_0 & G_0 \end{bmatrix} \begin{Bmatrix} V_2 \\ V_3 \\ V_4 \\ V_5 \\ V_6 \\ V_7 \\ V_8 \end{Bmatrix}, \qquad (3)$$

Решение этой системы уравнений с учетом, что стабилизированный ток задается через контакты №1 и №5 (т.е. $I_2 = I_3 = I_4 = I_6 = I_7 = I_8 = 0$) дает нам значения напряжений на контактах $V_1 = V_2 = V_3 = 0$ и $V_6 = V_7 = V_8 = V_5$. Так как ток, протекающий через контакт №5, равен $I_5 = -G_0 V_5$, выражения для продольного и холловского сопротивлений приобретают следующий вид:

$$R_{xx} = R_{15,24} = (V_2 - V_4)/I_5 = 0$$
$$R_{xy} = R_{15,37} = (V_3 - V_7)/I_5 = h/Me^2, \qquad (4)$$

демонстрируя целочисленный квантовый эффект Холла, который был впервые зарегистрирован Клаусом фон Клитцингом в 1980 году при исследовании магнетосопротивления двумерного электронного газа в кремниевых МОП-структурах в зависимости от напряжения на вертикальном затворе в условиях низких температур ($T$ = 1.5 К) и сильных магнитных полей ($B$ = 17.9 Тл) [32]. Было показано, что изменение холловского сопротивления на $h/e^2$ происходит каждый раз, когда уровень Ферми проходит через уровень Ландау. Таким образом, возможна регистрация ЦКЭХ как посредством приложения напряжения вертикального затвора, управляя плотностью двумерных носителей, при стабилизированном значении индукции магнитного поля, так и путём регистрации полевых зависимостей продольного и холловского сопротивлений при фиксированном положении уровня Ферми [33]. При этом справедливость выражения (4) доказана экспериментально с относительной точностью $10^{-7}$, что позволило успешно использовать КЭХ при создании эталона сопротивления [34]. Это стало возможным благодаря тому, что в сильных магнитных полях электронные состояния для носителей тока с различными по знаку проекциями волнового вектора («+*k*» и «-*k*» состояния) оказываются пространственно разнесены к противоположным краям образца [1]. Таким образом, ток переносится только в узком приповерхностном слое, и, несмотря на наличие дефектов и примесей, обратное рассеяние не происходит вследствие слабого перекрытия волновых функций для состояний с волновыми векторами «+*k*» и «-*k*». В результате реализуется баллистический режим транспорта носителей на макроскопических расстояниях (~1 мм в [33]), что приводит к регистрации нулевого продольного сопротивления мультиконтактного устройства, $R_{15,24} = 0$. Величина холловской разности потенциалов при этом будет определяться контактным сопротивлением, равным $h/Me^2$, где M – число краевых каналов.

Рассмотрим случай, когда матричные элементы, отвечающие состояниям «+*k*» и «-*k*» не эквиваленты: $G_{21} = G_{32} = G_{43} = G_{54} = G_{+k}$ и $G_{65} = G_{76} = G_{88} = G_{18} = G_{-k}$ (см. рис.2*а*). Путем алгебраических преобразований нетрудно показать, что в условиях стабилизации тянущего тока между контактами №1 и №5, измеренные значения продольного и холловско сопротивлений равны, соответственно, $R_{xx} = R_{15,24} = 0$ и $R_{xy} = R_{15,37} = 1/G_{+k}$. Иначе говоря, значение холловского сопротивления при стандартной геометрии эксперимента определяется только проводимостью канала «+*k*», даже при полном подавлении состояний «-*k*», $G_{-k} \rightarrow 0$. Продольная проводимость, измеренная 2-х контактным методом при использовании токовых контактов в качестве измерительных, также демонстрирует значение $G_{15,15} = 1/R_{15,15} = G_{+k}$, изменяясь на $e^2/h$ каждый раз, когда уровень Ферми проходит уровень Ландау. Падение напряжения при этом происходит не на всем протяжении исследуемой структуры, содержащей краевые баллистические каналы, а только на токовых контактах [16].

Подобное поведение проводимости было предсказано Ландауэром задолго до открытия ЦКЭХ, который показал, что если длина проводника становится меньше длины свободного пробега, $L_m$, и длины фазовой релаксации, $L_\varphi$, то реализуется баллистический режим транспорта носителей. Кондактанс проводника при этом оказывается равен $G = 2Me^2/h$, где $M = Int[2W/\lambda_F]$ соответствует целому числу фермиевских полуволн, $\lambda_F$, укладывающихся на его ширине, $W$ [35]. Таким образом, плавное уменьшение ширины проводника будет приводить к скачкообразному изменению кондактанса в единицах $e^2/h$, или т.н. «квантовой лестнице проводимости», что было подтверждено экспериментально в полупроводниковых одномерных каналах, полученных с помощью методики расщепленного затвора [26, 36-41].

Важным обстоятельством является возможность учета в рамках ФЛБ в матрице проводимости процессов обратного рассеяния в краевых каналах, а также наличие разупорядочения в контактах за счет присутствия дефектов и примесей [16]. В первом случае, если создаются условия, приводящие к возможности обратного рассеяния, например путем введения расщепленного затвора, в матрицу проводимости необходимо ввести новые элементы, ответственные за образовавшиеся связи между контактами с соответствующим весовым коэффициентом $p$, равным отношению числа каналов, участвующих в обратном рассеянии к полному числу краевых каналов. Допустим, в связи между контактами (8←7) и (3←2) стало участвовать не $M$, как ранее, а $N$ каналов. Тогда условие $G_{ij} = G_0$, больше не выполняется, и в матрице проводимости появятся члены $G_{28} = G_{73} = pG_0$, где $p = (M - N)/M$. При этом матричные элементы, ответственные за связи между контактами (8←7) и (3←2), на основании правил Кирхгофа также необходимо умножить на коэффициент $(1-p)$. Решение полученной системы уравнений приводит к следующему виду продольного и холловского сопротивлений:

$$R_L = h/e^2 \cdot [1/N - 1/M],$$
$$R_H = h/Me^2. \qquad (5)$$

Аналогичным образом можно учесть наличие разупорядочения в контактах за счет присутствия дефектов и примесей. В простейшем случае наличие дефекта вблизи одного из контактов приводит к уменьшению проводимости краевых каналов, связывающих его с соседними контактами, что в результате отражается на величине соответствующих матричных элементов. В частности, если в связи с разупорядоченным контактом участвует не $M$, а только один канал с проводимостью $e^2/h$, то холловское сопротивление равняется $R_H = h/M[1-p]e^2$ [16].

Наличие конечного сопротивления у контактных площадок также может существенно повлиять на вид измеряемой матрицы проводимости. Пусть для простоты все контакты и контактные площадки восьми контактного устройства идентичны друг другу и обладают сопротивлением $Rc$, и при этом в режиме ЦКЭХ реализуется только один краевой канал, $M = 1$. Тогда на основании правил Кирхгофа становится возможным рассчитать вклад сопротивлений $Rc$ в каждый из матричных элементов $G_{ij}$. На рисунке 2*b* представлены зависимости таких поправок к матричным элементам, связанным с контактом №1, от величины $Rc$. В силу симметрии задачи, остальные матричные элементы для прочих контактов будут претерпевать точно такие же изменения. Как видно из рисунка 2*b*, вклад сопротивлений контактных площадок остается несущественным до $Rc < 10$ Ом, после чего ошибка начинает нарастать, приходя в насыщение при значениях $Rc$, близких к значению проводимости одного краевого канала. Так, при $Rc > 5$ кОм идентифицировать краевой канал по виду матрицы проводимости становится затруднительным, так как поправка к матричному элементу, ответственному за связь между 1 и 8 контактами становится сравнима с его величиной. При этом измеряемые 4-х контактным методом продольное, $R_L$, и холловское, $R_H$, сопротивления так и останутся равными 0 и $h/e^2$, соответственно. Величина сопротивлений $Rc$ в этом случае скажется, прежде всего, на спектральной плотности и амплитуде шумов, сопутствующих измерениям [31].

## 2.3. Матрица проводимости в режиме квантового спинового эффекта Холла

Приведенный выше математический аппарат имеет большие перспективы для анализа транспортных свойств системы хиральных краевых каналов в топологических изоляторах при описании квантового спинового эффекта Холла (КСЭХ) [4]. В отличие от краевых каналов в режиме ЦКЭХ, топологические краевые состояния характеризуются наличием двух контуров, направленных в противоположные стороны для частиц со спином 1/2 и -1/2 в отсутствии внешнего магнитного поля (см. рис. 3*а*). Таким образом, при рассмотрении транспорта носителей в топологических изоляторах необходимо составить две матрицы проводимости для каждого из этих контуров, $G_\uparrow$ и $G_\downarrow$, аналогичные по своей структуре матрице $G$ в выражении (3). При этом направление движения в этом случае учитывается посредством транспонирования одной из матриц [11]. В результате, полный ток носителей в топологических изоляторах будет описываться в рамках формализма Ландауэра – Буттикера выражением (2) с матрицей проводимости $G_{КСЭХ} = G_{КЭХ} + G_{КЭХ}^T$. Рассчитанные таким образом сопротивления шести и четырех контактного мостика, $R = G_{КСЭХ}^{-1}$, демонстрируют значения: $R_{14,14} = 3h/2e^2$, $R_{14,23} = h/2e^2$, $R_{13,13} = 4h/3e^2$, $R_{13,54} = h/3e^2$ и $R_{14,14} = 3h/4e^2$, $R_{14,23} = h/4e^2$, соответственно, что хорошо согласуется с экспериментальными данными, полученными при изучении топологических изоляторов на основе гетероструктур HgTe/CdTe [11].

Тем не менее, правомерность применения операций транспонирования и последующего суммирования матриц проводимости для краевых каналов в режиме ЦКЭХ при описании спинозависимого транспорта требует дополнительно обоснования. Строго говоря, матрица вида $G_{КСЭХ}$ может быть получена только на основании правил Кирхгофа в предположении, что между двумя соседними контактами имеется двусторонняя связь с проводимостью $G_0=e^2/h$, как в случае реализации баллистического канала, носители в котором полностью поляризованы по спину, а рассеяние назад возможно только на контактах. Поэтому при детальном рассмотрении матричные элементы $G_{ij}$ и $G_{ji}$, отвечающие различным направлениям движения и спина, должны вычисляться с учетом спиновой поляризации носителей и процессов спинозависимого рассеяния. Примером влияния спиновой поляризации носителей на величину проводимости одномерного канала является регистрация расщепления первой ступеньки квантовой лестницы проводимости вблизи значения $0.7(2e^2/h)$ [26, 41-44]. Это расщепление, получившее название «0.7-особенности», при приложении внешнего магнитного поля трансформируется в ступеньку с амплитудой $G = 0.5(2e^2/h)$, что соответствует полному снятию спинового вырождения для носителей в канале [42]. При этом уменьшение плотности носителей посредством приложения напряжения вертикального затвора, также приводит к аналогичной трансформации «0.7-особенности», вследствие процессов спонтанной спиновой поляризации в нулевом магнитном поле [19, 26, 45].

Формализм Ландауэра-Буттикера позволяет учесть неэквивалентность в проводимости краевых каналов для носителей с различными спинами. Предположим, что $G_{j,j-1} = G_\uparrow$ и $G_{j,j+1} = G_\downarrow$, где $j$ циклически может принимать значения от 1 до 8 (см. рис.3*а*). Можно показать, что в этом случае для продольного и поперечного сопротивлений восьми контактного холловского мостика выполняются следующие соотношения:

$$R_{xx} = R_{15,24} = G_\uparrow \cdot G_\downarrow \cdot (G_\uparrow + G_\downarrow)/(G_\uparrow^4 + G_\downarrow^4)$$
$$R_{xy} = R_{15,37} = (G_\uparrow - G_\downarrow) \cdot (G_\uparrow + G_\downarrow)^2/(G_\uparrow^4 + G_\downarrow^4). \qquad (6)$$

Таким образом, наличие рассогласования между проводимостями краевых каналов для носителей с противоположными проекциями спина, $G_\uparrow \neq G_\downarrow$, приводит к уменьшению величины измеряемого сопротивления, что наблюдалось экспериментально при изучении нелокального транспорта в топологических изоляторах на базе HgTe/CdTe [11]. При этом наиболее существенное уменьшение сопротивления было зарегистрировано для *H*-образного мостика, при использовании 2-х контактного метода измерений, $R_{1414}$, особенностью которого, напротив, является превышение истинного значения сопротивления объекта исследования за счет наличия конечных сопротивлений у контактов, $R_{2K} = R_{структуры} + R_{конт «исток»} + R_{конт.«сток»}$. Учет неэквивалентности в проводимости краевых каналов для носителей с различными спинами согласно выражениям (6) позволяет снять это противоречие, а также объяснить возникновение ненулевого холловского сопротивления в режиме КСЭХ, наблюдавшееся ранее при изучении кремниевых наносандвичей, представляющих собой кремниевые квантовые ямы *p*-типа с плотностью двумерных дырок $p_{2D} \sim 1.1 \cdot 10^{14}$ м$^{-2}$, ограниченные δ-барьерами, сильно легированными бором [7]. Зарегистрированные величины тензора проводимости кремниевых наносандвичей, выполненных в геометрии восьми контактного холловского мостика, демонстрировали значения $G_{xx} = R_{xx}/(R_{xx}^2+R_{xy}^2) = 4e^2/h$, и $G_{xy} = R_{xy}/(R_{xx}^2+R_{xy}^2) = e^2/h$, которые также могут быть объяснены в рамках описанного выше подхода в предположении $G_\uparrow = 4.395 \cdot e^2/h$ и $G_\downarrow = 3.88 \cdot e^2/h$. При этом полученная величина проводимости одного канала, $G \sim 4 \cdot e^2/h$, неоднократно наблюдалась при изучении квантового транспорта в низкоразмерных структурах, выполненных на базе различных материалов. В частности, в графене при регистрации КЭХ холловская проводимость претерпевает изменения, кратные $4 \cdot e^2/h$, что обусловлено наличием спинового и долинного вырождения в дираковской точке зонной структуры графена [46]. В топологических изоляторах на основе гетероструктур HgTe/CdTe при исследовании *H*-образных структур четырех контактным методом, проводимость $G_{14,23} = 4 \cdot e^2/h$ возникала вследствие наличия разнонаправленных краевых каналов для носителей с противоположными спинами [11]. Точно такая же величина проводимости предсказывалась для процессов, сопровождающихся передачей удвоенного заряда 2*e*, а именно многократного Андреевского отражения и парного туннелирования [47].

Как и в случае ЦКЭХ, наличие у контактных площадок конечного сопротивления, *Rc*, приводит к весьма существенным поправкам к измеряемым элементам матрицы проводимости. Так, ошибка в измерении элементов, отвечающих проводимости краевых каналов, уже начиная с $Rc > 100$ Ом, превышает 1% (см. рис. 3*b*). При этом в отличие от ЦКЭХ, в режиме КСЭХ матрица проводимости обладает более высокой степенью симметрии, обусловленной требованием выполнения симметрии по отношению к обращению времени в отсутствии внешнего магнитного поля, что приводит к уменьшению количества различных матричных элементов.

## 2.4. Учет влияния наличия сопротивления контактных площадок при построении матрицы проводимости

Влияние наличия у контактных площадок конечного сопротивления на величины элементов матрицы проводимости может быть минимизировано. Для этого при использовании трех контактной схемы измерения разности потенциалов необходимо вычесть из измеренной величины $U_{ij}$ падение напряжения $Uc_i = Rc_i \cdot I_{ik}$, где $I_{ik}$ – ток, стабилизированный между *i*-ым и *k*-ым контактами. Значения сопротивлений $Rc_i$ в дрейфовом режиме могут быть оценены посредством сопоставления данных четырех- и двух- контактного методов измерения сопротивления исследуемого мультиконтактного устройства с учетом его удельной проводимости и геометрических размеров. Данные оценки являются очень грубыми, поскольку не учитывают эффектов растекания тока $I_{ik}$ по образцу прямоугольной формы при различном выборе *i*-ого и *k*-ого контактов. Однако даже такая точность в определении $Rc_i$ позволяет существенно уменьшить ошибку в определении матрицы проводимости. Как показывает анализ поведения соответствующих поправок к матричным элементам $G_{ij}$, учет сопротивлений $Rc_i$ с точностью хотя бы до 10 Ом сводит ошибку $\Delta G_{ij}$ к величине < 0.1% (см. рис. 2*b*, 3*b*). В случае КЭХ и КСЭХ величины контактных сопротивлений также могут быть оценены описанным выше способом. Для этого необходимо перевести транспорт в исследуемом образце из баллистического в дрейфовый режим, например, посредством снятия магнитного поля в режиме КЭХ.

Существенная зависимость вида матрицы проводимости от величины сопротивлений контактов может быть использована при разработке аналоговых устройств криптографии с закрытым ключом, в качестве которого выступает 8-и символьный «пароль», состоящий из $Rc_i$, где $i = 1 – 8$. В этом случае информация шифруется одним из классических симметричных криптографических алгоритмов в матричной форме, размерности 8 x 8, после чего создается восьми контактное устройство с матрицей проводимости, эквивалентной матрице с зашифрованной информацией, и производится повторное шифрование посредством введения дополнительных контактных сопротивлений, выступающих в качестве «пароля».

Для того чтобы продемонстрировать принцип действия предложенного алгоритма шифрования, рассмотрим конкретный пример. Предположим, мы хотим зашифровать сообщение «ALL_YOU_NEED_IS_LOVE. _LENNON», и сохранить его в аналоговом криптографическом устройстве. Для этого каждой букве сообщения ставится в соответствие число, равное ее порядковому номеру в английском алфавите, пробелу – 27, точке – 28. Сообщение приобретает вид (1), (12), (12), (27), (25), (15), (21), (27), (14), (5), (5), (4), (27), (9), (19), (27), (12), (15), (22), (5), (28), (27), (12), (5), (14), (14), (15), (14). Из полученного набора чисел построчно составляется верхняя треугольная матрица размером 8 x 8 с нулевыми диагональными элементами, $\mathbf{G_\blacktriangledown}$. Далее посредством сложения полученной матрицы $\mathbf{G_\blacktriangledown}$ с ее эрмитовым сопряжением, $\mathbf{G_\blacktriangledown} + \mathbf{G_\blacktriangledown}^\mathrm{T}$, и последующего вычисления диагональных элементов по формуле $G_{ii} = -\sum_j G_{ij}$, становится возможным получить матрицу проводимости криптографического устройства, содержащего зашифрованное послание. Однако такой шифр является весьма уязвимым к взлому. Имея достаточный объем зашифрованной информации, его можно вскрыть посредством сопоставления частот повторения символов и частот букв, используемых в английском языке. Поэтому проводится повторное шифрование путем последовательного введения дополнительных сопротивлений к контактам 1 – 8, с номиналами, равными в Омах: (2), (5), (1), (20), (12), (5), (19), (27), что согласно используемому шифру соответствует паролю «BEATLES_». В результате сообщение, содержащееся в матрице проводимости нового устройства, принимает вид: (1.766), (12.715), (11.947), (26.108), (24.926), (14.556), (18.868), (27.03), (13.395), (5.293), (5.321), (4.324), (23.697), (9.065), (18.53), (26.855), (11.701), (13.704), (19.712), (5.219), (24.458), (22.206), (11.604), (5.106), (12.245), (13.204), (12.42), (12.085). Частотные методы анализа в этом случае не дадут результата, так как один и тот же символ может ассоциироваться с различными числами. Сложность шифра можно существенно повысить посредством увеличения номинала сопротивлений контактов, однако это приведет к необходимости учитывать больше значащих цифр после запятой при определении элементов матрицы проводимости. В частности, при том же самом пароле, задаваемом сопротивлениями с номиналом в $10^3$ Ом, для корректного дешифрования сообщения необходимо измерить элементы матрицы проводимости с относительной точностью, превышающей $10^{-6}$.

Тем не менее, создание мультиконтактного устройства с заданной матрицей проводимости является весьма сложной технической задачей. Однако её можно значительно упростить, если перейти к двоичной системе кодирования. В этом случае информацию в матрице проводимости будет содержать лишь последовательность нулевых и ненулевых элементов, отвечающих отсутствию или наличию связи между соответствующими контактами. На рисунке 4 приведён пример принципиальной схемы такого устройства, которое может быть получено в рамках стандартной планарной технологии, для заданной матрицы с размерностью 4 x 4.

### 3. Заключение

1) Предложен метод измерения матрицы проводимости мульти контактных полупроводниковых структур. В основе метода лежит решение СЛАУ на базе уравнений Кирхгофа для каждого из контактов, составленных из разностей потенциалов $U_{ij}$, измеренных при стабилизации токов $I_{kl}$, где $i,j,k,l$-номера контактов. Полученная матрица полностью описывает исследуемый объект, отражая его геометрию и однородность. Данный метод может найти широкое применение при использовании формализма Ландауэра-Буттикера для анализа транспорта носителей в режимах КЭХ и КСЭХ. Отдельный интерес представляет исследование полевых зависимостей элементов матрицы проводимости в диапазоне магнитных полей, отвечающем переходу между соседними плато в режиме ЦКЭХ.

2) Для продольного и холловского сопротивлений восьми контактного устройства в режиме КСЭХ получены аналитические выражения, учитывающие наличие неэквивалентности краевых каналов для носителей с противоположными проекциями спина. Предложенный подход позволяет качественно объяснить, почему при использовании 2-х контактного метода измерений наблюдается уменьшение продольного сопротивления относительно теоретически предсказанного значения, а также – почему возникает ненулевое холловское сопротивление в исследованиях двумерных топологических изоляторов.

3) Детально исследован вклад сопротивления контактных площадок, $Rc$, в формирование элементов матрицы проводимости в режимах КЭХ и КСЭХ. Показано, что, начиная с $Rc>100$ Ом, величина сопротивления контактных площадок оказывает существенное влияние на вид матрицы проводимости. Предложен метод по минимизации этого влияния.

4) Продемонстрированы возможности практического применения полученных результатов при разработке аналоговых криптографических мультиконтактных устройств с закрытым ключом типа $N$-символьного «пароля», каждый символ которого ассоциируется с величиной сопротивления одного из $N$ контактов.

Тексты программ по расчету элементов матрицы проводимости и их погрешностей доступны в интернете:
https://drive.google.com/#folders/0B7uvzmnf5CowYUtpU1REWFdLM_1U

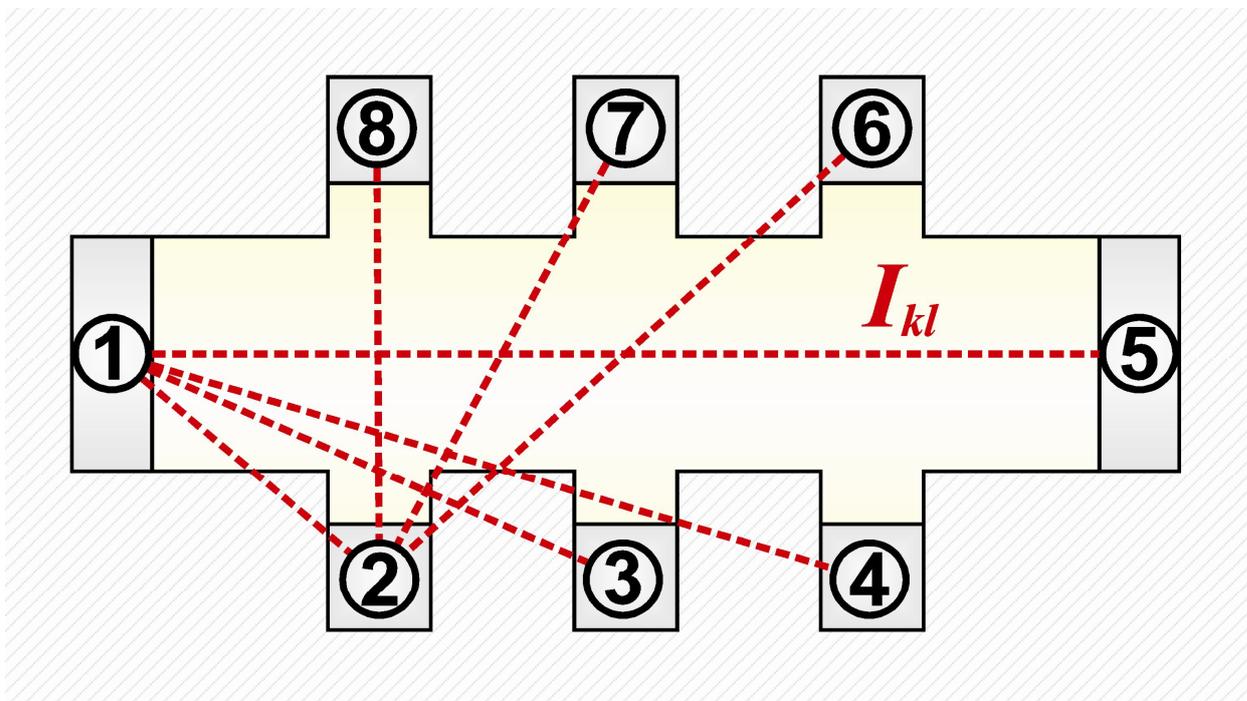

**Рис. 1.** Схема метода регистрации матрицы проводимости для восьми контактного холловского мостика. В основе метода лежит решение систем линейных алгебраических уравнений на базе уравнений Кирхгофа для каждого из контактов, составленных из разностей потенциалов $U_{ij}$, измеренных при стабилизации токов $I_{kl}$, где $i,j,k,l$-номера контактов. При этом токовые контакты необходимо выбирать так, чтобы среди набора не было двух пар, эквивалентных с точки зрения симметрии, $(k,l)$ = (1,5); (1,4); (1,3); (1,2); (2,6); (2,7); (2,8).

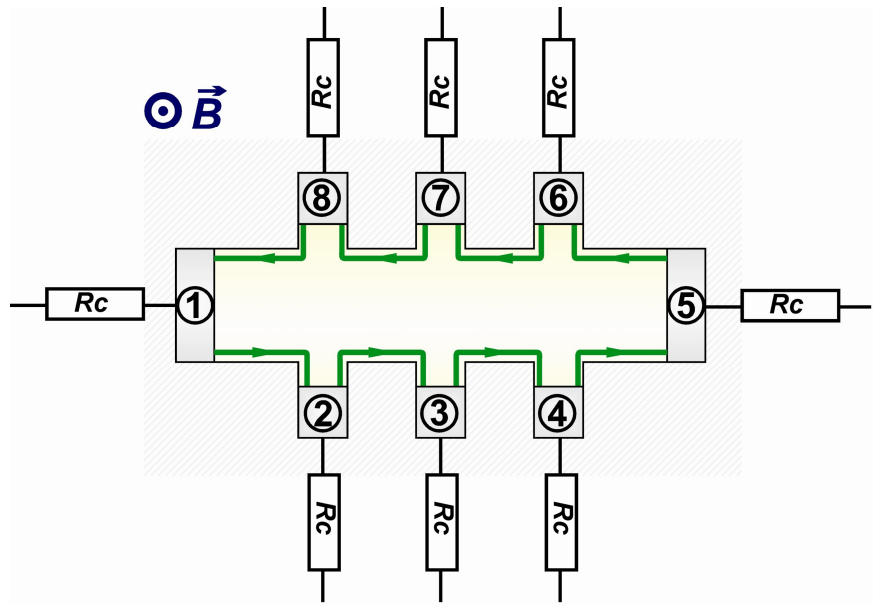

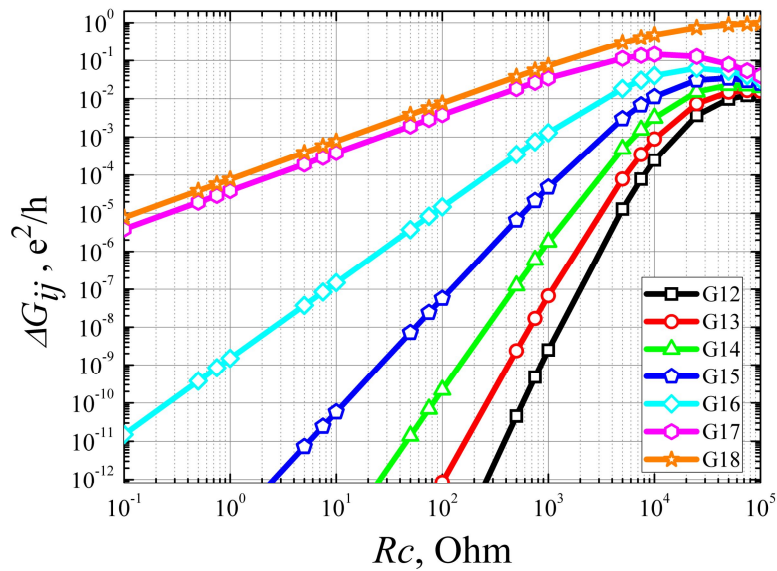

**Рис. 2.** *а* – схема краевых каналов, возникающих в восьми контактном холловском мостике в режиме ЦКЭХ. Каждый краевой канал, соединяющий *i* и *i+1* контакты, соответствует матричному элементу $G_{i+1,i}=M \cdot e^2/h$, матрицы проводимости, *M* – целое число. Диагональные элементы матрицы при этом равняются $G_{ii}=-G_{i+1,i}$, все прочие матричные элементы равны нулю, что обусловлено требованием выполнения правил Кирхгофа.

*b* - Зависимости поправок к матричным элементам $G_{1j}$, обусловленных наличием у контактных площадок конечного сопротивления, *Rc*, от его величины в случае *M*=1, $Rc_i = Rc$ для *i* = 1 - 8. При *Rc* = 0 ненулевым является только элемент $G_{18}$ (поправка к нему представлена кривой, маркированной звездой). В силу симметрии задачи, прочие матричные элементы будут претерпевать аналогичные изменения.

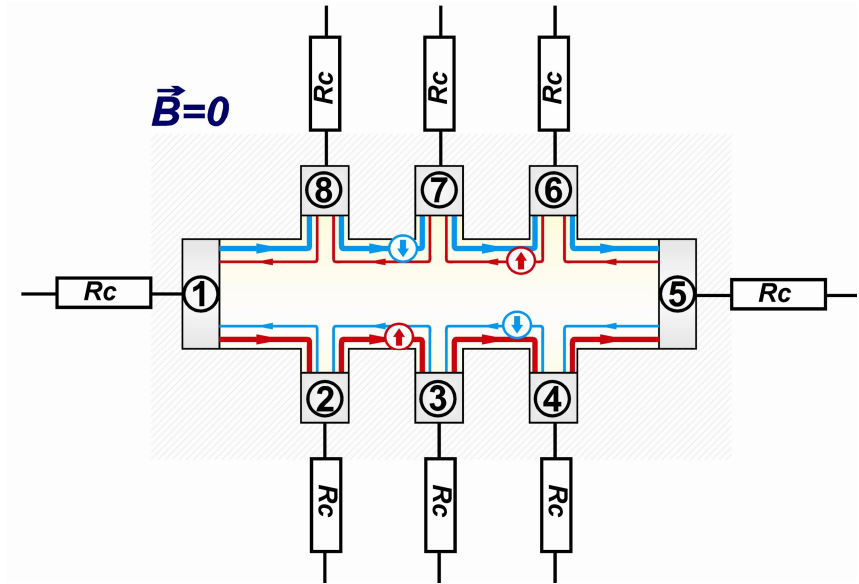

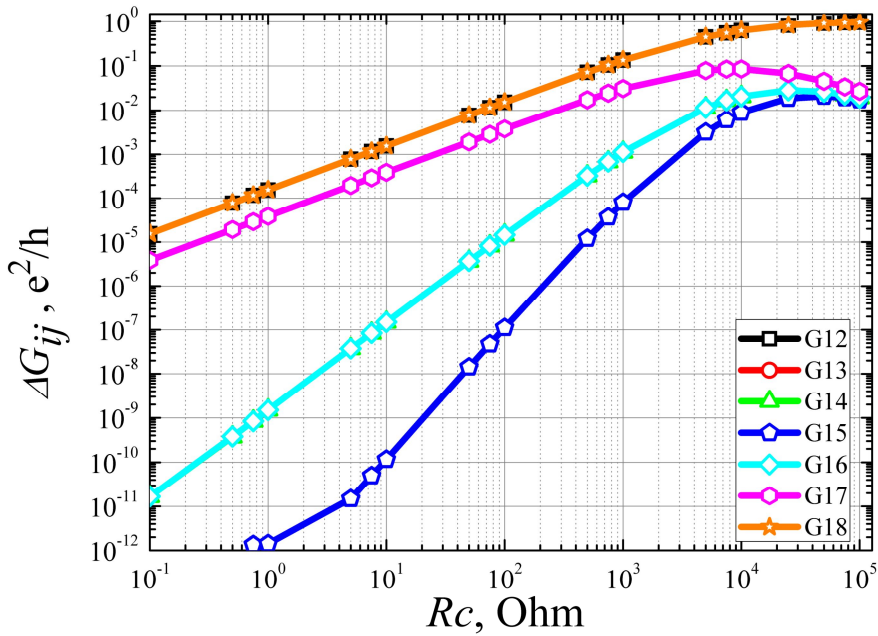

**Рис. 3.** *a* – Схема краевых каналов, возникающих в восьми контактном холловском мостике в режиме КСЭХ. Вертикальными стрелками отмечен знак проекции спина носителей в краевых каналах. *b* - Зависимости поправок к матричным элементам $G_{1j}$, обусловленных наличием конечного сопротивления контактных площадок, $Rc$, от его величины для случая, представленного на рис. 3*a*. При $Rc = 0$ ненулевыми являются только элементы $G_{18} = G_{12} = e^2/h$ (поправки к ним представлены кривыми, маркированными звездой и квадратом, соответственно).

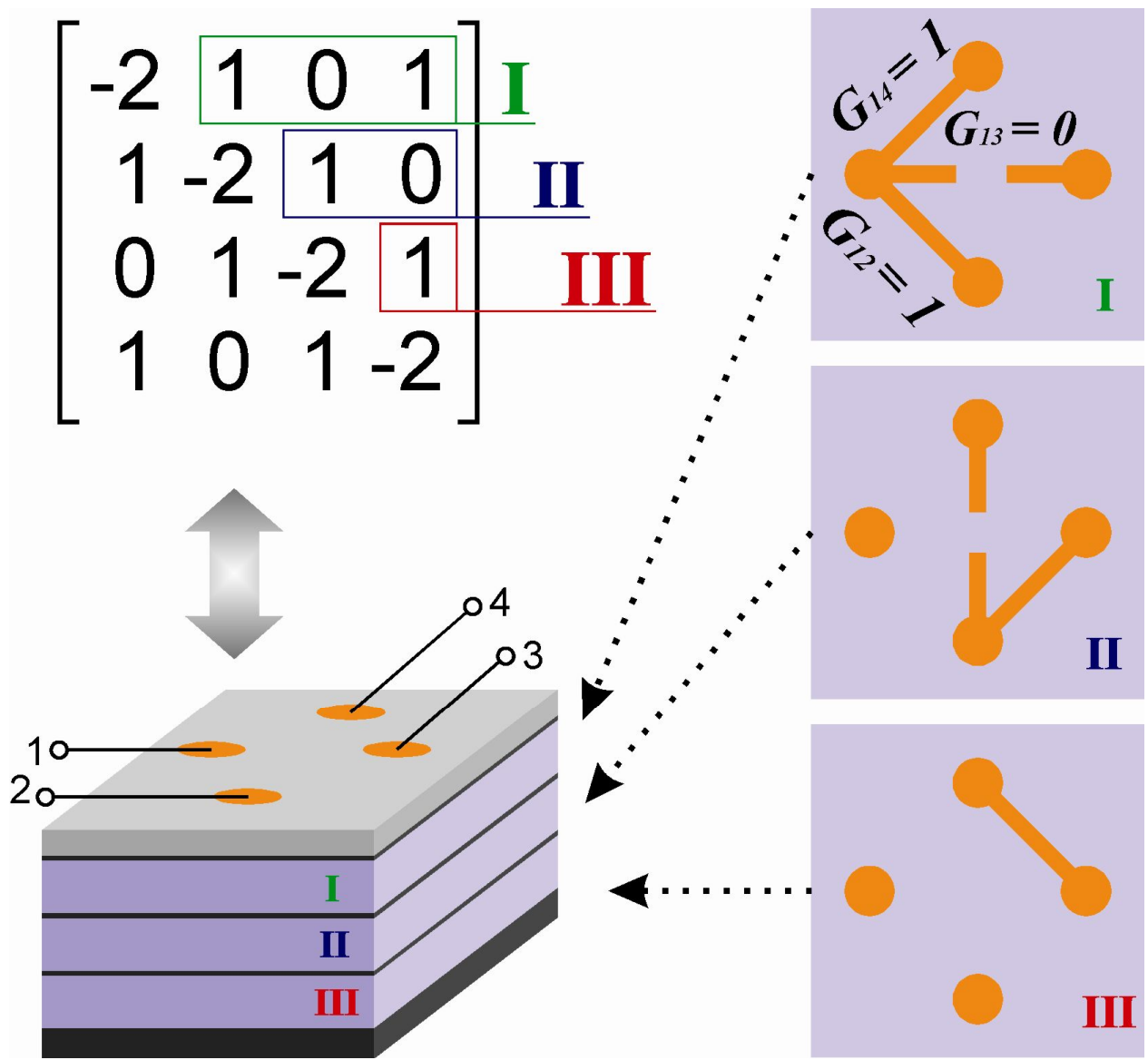

**Рис.4.** Принципиальная схема создания четырех контактного криптографического устройства, ассоциирующегося с заданной матрицей проводимости размерности 4 x 4, информация в которой кодируется последовательностью нулевых и ненулевых элементов (в данном примере: 101101), отвечающих отсутствию ($G_{ij} = 0$, разорванная черта на рисунке) или наличию связи между соответствующими контактами ($G_{ij} = 1$, прямая черта). Устройство создаётся послойно, каждый из слоев (I – III в данном примере) отвечает строке в матрице проводимости с учетом требования $G_{ij} = G_{ji}$. Диагональные элементы при этом равняются сумме всех элементов в строке, взятой со знаком «-».